\documentclass[namedreferences,hyperref,optionalrh,solaromanenum]{spr-sola}

\usepackage{graphicx}                    
\usepackage{color}   

\newcommand{\aap}{{\it Astron. Astrophys.}}

\newcommand{\apj}{{\it Astrophys. J.}}

\chardef\us=`\_

\newcommand\dblquote[1]{\textquotedblleft #1\textquotedblright}
\newcommand\sglquote[1]{\textquoteleft #1\textquoteright}

\begin{document}

\begin{frontmatter}

\title{The Current State of the Controversy over Screening in Nuclear Reactions}

%
\author[addressref={aff1},email={dappen@usc.edu}]{\inits{W.}\fnm{Werner}~\snm{D\"appen}\orcid{0009-0008-5933-8977}}
\address[id=aff1]{Department of Physics and Astronomy, USC, Los Angeles, CA 90089-0484, U.S.A.}
%

\begin{abstract}
A controversy about the possibility of dynamic effects in nuclear screening has been around for several decades. On the one hand, there is the claim that there are no dynamic effects, and that the classic Salpeter correction based on static Debye screening is all that is needed for astrophysical applications. The size of the correction is on the order of 5\% in typical solar fusion reactions. On the other hand, numerical simulations have shown that there is a dynamical effect, which essentially cancels the Salpeter correction. The results of the numerical simulations were later independently confirmed. The astrophysical community, however, has so far largely ignored the possibility of dynamical screening. The present paper is meant to serve as a reminder of the controversy. Not only does the claim of an absence of a dynamical effect equally warrant an independent confirmation, but there is motivation for further investigation, such as the assessment of current laboratory experiments and a quantitative study of the dynamical effect in case it will turn out to be real.
\end{abstract}

%

\end{frontmatter}

%
\section{Introduction}
     \label{S-Intro} 
     
It is well known that ordinary stars can only operate thanks to the possibility of quantum tunneling. Before nuclei undergo reactions they have to overcome the Coulomb barrier which is on the order of MeVs. In stable phases of stellar nuclear burning, such as during solar main-sequence life, core temperatures are on the order of keVs. By standard Maxwell-Boltzmann probability distribution, the chance of a pair of nuclei to penetrate the Coulomb barrier is therefore on the order of $e^{-1000}$, which is tantamount to impossible. Quantum tunneling gave the solution of this paradox. The first person to apply the Schrödinger equation to a problem which involved tunneling between two classically allowed regions through a potential barrier was Friedrich Hund
(\citeyear{hund27}). Shortly afterwards,~\cite{gam29} computed the tunneling rate in nuclear decay, 
while Atkinson and Houtermans~(\citeyear{atk29}) applied Gamov's result to solar fusion. The Maxwell-Boltzmann distribution still plays an important role but it is now a factor multiplied with the tunneling probability and 
with the genuine reaction rate from nuclear physics, usually assumed to be independent of the other effects. Underlying this factorizing is the \emph{quasi-classical assumption} that in a random distribution, the tunneling probability for a given pair of colliding nuclei with a certain relative velocity at infinite separation,  is the same as that for a coherent stream of that relative velocity. The product of the two probabilities is the so-called \sglquote{Gamov peak}, which typically lies at temperatures of 3--4\, $kT$. 

The seminal discussion of the the relevant nuclear reaction was made by \cite{bet39}, but until the 1950s it remained an open question, if the main contribution to solar energy came from the pp chain or the CNO cycle. (Regarding the history of the CNO cycle in this context, see~\citealp{wie18}.) The dominant role of the pp chain
was established by Salpeter (\citeyear{salp52},\citeyear{salp53}). Soon after, \cite{salp54} introduced a modification of the tunneling probability. Arguing that two colliding nuclei are not isolated from their surroundings but that their charges are rather screened in the overall neutral ionized plasma, he computed the tunneling probability not with the bare Coulomb potential but the Debye-H\"uckel~(\citeyear{DH23}) potential. Since this is lower than the Coulomb potential, the tunneling probability becomes higher, a result commonly referred to as the so-called \sglquote{Salpeter screening enhancement}.

Salpeter assumed that one can still multiply the Maxwell-Boltzmann distribution of the nuclei with the tunneling probability, in the same way as had been done with the bare Coulomb potential. The Debye-H\"uckel potential he considered is time-independent, since the underlying theory is thermodynamic. Therefore, Salpeter screening is also inherently static. However, the problem with static screening is that, intuitively, the height of the screened potential must depend on the velocity of the colliding nuclei. This is illustrated by a pair of very rapidly colliding nuclei. One would expect that the cloud of the screening electrons can still easily adjust during the flight of the nuclei, but the surrounding positive charges would be frozen at the same time, therefore precluding the attainment of the overall re-adjusted screening potential. However, precisely such velocity-dependent effects are not taken into account in the Salpeter theory. 

However, with a velocity-dependent tunneling probability, the standard assumption that one can multiply the probabilities, that of finding a pair of nuclei with a certain relative velocity, and that of a particle of that velocity tunneling through the Coulomb barrier. In the absence of screening, this is a reasonable assumption, which simply states the independence of two events. However, when the barrier itself becomes velocity dependent, the simple product rule breaks down. If the overall reaction rate, that is, the Gamov peak, were at the equilibrium temperature $kT$, the problem might be less severe, since around $kT$ one could imagine that velocity-dependent effects essentially cancel. However, the Gamov peak is at 3-4 $kT$, which selects those velocities that are most relevant for nuclear energy production. Therefore, it is by far not obvious if one can still use the original Debye-H\"uckel potential for these fast nuclei.

Of course, even then it cannot be excluded that in the net nuclear reaction rate, all deviations from the static result will somehow cancel out. However that would have to be demonstrated. Effects from velocity-dependent tunneling probabilities (resulting from a velocity-dependent \sglquote{mountain}) are generally referred to as \sglquote{dynamic screening}, which was first considered several decades ago~\citep{mit77,car88}. A detailed history of dynamic screening is given in Appendix~B of \cite{sha00}, while~\cite{dz95} discuss the impact of various screening formalisms for stars (in a study that was then done in the context of finding a possible solution to the solar-neutrino problem).

One approach to demonstrate the possibility of dynamic screening was pioneered by applying molecular-dynamic (MD) simulations of a large number of electrons and nuclei \citep{sha96}. Another was to attempt to compute the overall reaction rate with the methods of quantum statistical mechanics \citep{bs97}. Their respective results disagree significantly. In the following I will present these different approaches and the ensuing controversy. Then I will discuss the astrophysical relevance of the controversy and end with a plea for necessary future work.

\section{Salpeter's Static Screening} 
      \label{SalpSS}
      
\cite{salp54} derived an expression for the enhancement of
nuclear reaction rates due to electron screening. When the Debye-H\"uckel theory
of dilute solutions of electrolytes is applied to electrons and ions in a plasma under the
condition of weak screening ($\phi_{\rm{interaction}}\ll k_{\rm B}T$), the interaction
potential of a charge $Z_ie$ becomes the static-screened Coulomb potential (for simplicity,
here we only consider
the case of one ion species)
\begin{equation}
\phi(r) = {Z_ie\over r}e^{-r/R_{\rm D}} \ ,
\label{E.3.11}
\end{equation}
with the Debye length $R_{\rm D}$ given by
\begin{equation}
R_{\rm D} = \sqrt{k_{\rm B}T\over{ 4\pi\left({n_e e^2 + n_i Z_i^2e^2}\right) } } \ .
\label{E.3.12}
\end{equation}
where $n_e,n_i$ are electron and ion densities, respectively.

The outcome of screening is conveniently expressed by the \sglquote{enhancement factor} $f$, a multiplier of the unscreened result to yield the screened reaction rate. Under solar central conditions, a typical correction is about $f$=1.05~(see Figure 2 of \citealp{we00}), while $f$=1 corresponds to the absence of screening effects.

\section{Beyond Salpeter} 
      \label{Beysal1}
      
\subsection{Shaviv and Shaviv's Molecular-dynamic Simulations} 
  \label{SSMD}
  
\citeauthor{sha01} (\citeyear{sha96},\citeyear{sha00},\citeyear{sha01}) applied MD techniques
to follow the motions of particles under conditions found in the solar center. They simulated the detailed motion of 
a large number of ions and electrons subject to Coulomb interactions. Electrons were treated by a quasi-classical approximation whose idea goes back to G\"unter Kelbg (\citealp{ke63}; for the history and impact of
those approximations, see the recent review by \citealp{bo23}). See \cite{sha96} for the specific quasi-classical approximation used in this MD calculation. 

By following the local motion of each particle, the mean-field assumption underlying Salpeter's static screening was avoided. Because of the vastly different mass of the electrons and the nuclei, such simulations are challenging, much harder than those of gravitationally interacting assemblies of stars. While the latter simulations can done with millions or even billions of particles, Shaviv and Shaviv's calculations could only involve thousands.

By calculating the pair correlation functions of their simulation, Shaviv and Shaviv found that they are substantially different from those obtained by the formalism of Salpeter. And the resulting enhancement factor became essentially~1, expressing the fact that the dynamical effects practically obliterate the Salpeter screening. The most recent summary of that group's work is \cite{sha10}. 

\subsection{Independent Confirmation of Shaviv and Shaviv's Results} 
  \label{USCMD}

\begin{figure}[ht]
\centerline{\includegraphics[width=0.9\textwidth,clip=]{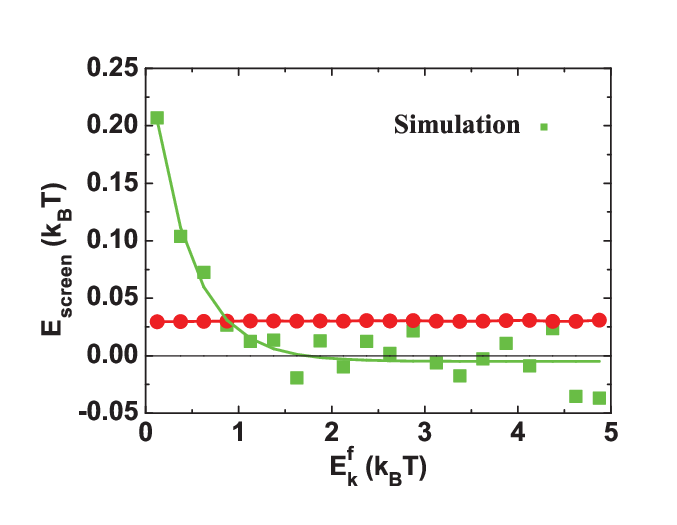}}
\caption{Dynamic screening energy at the turning point for pairs of
           protons with a given relative kinetic energy. For comparison, the
           static screening energy evaluated at the average turning point of proton
           pairs with each energy is also shown (data from~\citealp{mao09}).}
\label{Fig1}
\end{figure}

The results of Shaviv and Shaviv were later independently confirmed at USC by my former graduate students (\citealp{mao09,mu11,dm12}). We essentially performed the same molecular dynamic calculation as the Shavivs', using 1000~protons
and electrons, and obtained virtually the same results (see Figure~\ref{Fig1} and Table~\ref{tab:integrals}). If anything, the new results hinted even at a tiny reduction of the unscreened result.
While I consider this reduction insignificant, it has nonetheless received attention \citep{jcd21}.

\begin{table*}[h]
 \begin{center}
  \begin{tabular}{|l|c|c|}
    \hline
     Case & Screening energy U & Reaction-rate correction   \\ \hline \hline
     Unscreened  & 0  & 1        \\ \hline
     Statically screened & $U_0 = -{Z_1Z_2e^2}/{R_D}$  & 1.042   \\ \hline
     Dynamically screened  & $U_0(E)  = k_{\rm B}T\cdot   $ & \\
                           & $\left( 0.005 - 0.281\;{\rm{exp}}\left(-2.35 {E}/{k_{\rm B}T}\right)\right)$ & 0.996 \\ \hline
   \end{tabular}
  \end{center}
 \caption{Screening energies and the ratio of screened to unscreened nuclear reaction rates for solar p-p reactions as a function of the relative kinetic energy $E$ of the colliding protons. The dynamical result is a fit to numerical data (from~\citealp{dm12}).}
 \label{tab:integrals}
\end{table*}

\subsection{Brown and Sawyer's Analytical Results} 
      \label{Beysal2}
      
\cite{bs97} endeavored a rigorous \emph{ab-initio} computation of the plasma effects
involved in nuclear fusion.
The authors pointed out that in the so-called \sglquote{basically classical} approach, 
there are conceptual problems raised by the division of the problem into 
a quantum-mechanical and a classical part. They asserted that the 
\dblquote{literature lacks any development that begins with a correct 
general expression for the rate}.
The key ideas in their work were (i), treating the screening correction as a quantum observable, 
and (ii), proceeding through so-called \sglquote{imaginary time} expansions, thus allowing that the 
statistical-mechanical problem can be mapped into a quantum dynamical problem.
At the end of an elaborate 26-page analysis, they come to the conclusion that for solar
nuclear fusion at least, their result
essentially reproduces the result of~\cite{salp54}. The authors then conclude further
that there are no dynamical effects under the given conditions. Finally, \citeauthor{bs97} 
conjecture that any approach not agreeing with their results \dblquote{will miss physics 
that is as important as the physics that it includes}.

\section{Screening Enhancement in the Astrophysical Community} 
      \label{Current}

The late John Bahcall, who pioneered solar standard models and neutrino
flux predictions~\citep{bah82}, had a decisive influence on the astrophysical community.
The static screening result of~\cite{salp54} is now broadly accepted. After teaming up with Brown and Sawyer, Bahcall and his collaborators~\citep{bah02} consistently and tirelessly propagated the Salpeter result,
which they believed as certified by an exact
analytical calculation. The same recommendation is given by two comprehensive papers on solar fusion,
each authored by a large group \citep{ad98,ad11}. The first of these papers mentions the screening controversy, but it still basically sides with~\cite{bs97}. The second notes at least that
\dblquote{the controversy has not completely died down}, citing the USC results. A very recent review by~\cite{ali22} does not directly
take sides, mentioning both Shaviv and Shaviv's simulations and their independent confirmation, but then
goes on saying that their outcome \dblquote{has, however, been disputed by Bahcall and collaborators who argued that Salpeter’s screening approach is valid also at the Gamow peak energy due to the nearly perfect thermodynamic equilibrium present in the solar plasma}. However, the existence of a nearly perfect thermodynamic equilibrium does not necessarily preclude dynamical effects at the high-velocity end of the Maxwell-Boltzmann distribution. Such effects could take place without interfering with the overall thermodynamic equilibrium. Therefore, the real reason that~\cite{bah02} had ruled out significant deviations from the Salpeter result is their faith in the result of~\cite{bs97}.

Other authors ignore the controversy altogether. \cite{bjcd22} use the nuclear reaction rates of \cite{ad11}, which is adopting Salpeter screening for solar conditions. And \cite{lio00} comes to the conclusion that in screening \dblquote{nonlinear effects are shown to be negligible}, meaning that going to higher-order terms within Salperter's theory is not necessary. This is true, but it doesn't address the controversy. The controversy is about abolishing the Salpeter correction altogether. I am aware of only one suggestion to take dynamical screening seriously,
and that is coming from outside stellar and solar physics, namely cosmology. \cite{hwa21} discuss dynamical screening effects on big-bang nucleosynthesis. While they say that \dblquote{our result shows that the dynamical screening effects do not affect the primordial abundances} they do not exclude finer astrophysical effects.
In particular they mention that \dblquote{the dynamical screening effects on the CNO cycle are more effective than that on the p-p chain}, which could provide a \dblquote{correction of the CNO cycle for the solar evolution as well as CNO neutrino detection}. Therefore they conclude that \dblquote{if the dynamical screening effects are visible under the solar condition, those effects leave several issues worth discussing for related plasma properties in other astrophysical environments}.

Nevertheless, by and large, the stellar and solar communities continue to ignore the findings from molecular dynamics. Nobody seriously proposes to give equal consideration, both to the Salpeter screening enhancement and to a
no-screening result. And while \cite{sha10} is obviously happy to mention the independent confirmation of his 
group's MD simulations, so far there hasn't been a call to repeat the study of \cite{bs97}.

\section{Conclusion} 
\label{S-Conclusion} 

Given the undisputable role of \cite{bs97} in the screening controversy, an effort should be made to repeat their analysis. Also, further progress about the controversy could come from laboratory experiments (\citealp{cas23}, \citealp{wu17}). An entirely different approach to study dynamic screening in nuclear fusion was proposed by \cite{and10}, who have used an apparent \sglquote{duality} to infer the solar screening results from a system under totally different physical conditions, in their case experiments on a laser-cooled ionic system. 
Furthermore, it might be worth to mention that there are alternative simulation techniques in other areas of plasma physics. For instance, people have been using so-called \sglquote{particle-in-cell (PIC) simulations}, pioneered by~\cite{lan70}, to study dynamic effects on scales that include the Debye length, albeit in a much more rarefied plasma~(a typical example is \citealp{rek14}). In particular, these approaches treat the dynamics of electrons more seriously than the MD simulations of dynamic screening. Therefore, studying dynamic screening with PIC techniques could be a worthwhile extension of the MD approach.

And, last but not least, the astrophysical detectability of the screening effect should be re-visited in detail, even if \cite{we00} claim that the \sglquote{no-screening} case is ruled out by present-day helioseismic data. Their claim is based on a helioseismic sound speed inference in the region of $0.2-0.8$R$_{\odot}$, with an assumed \sglquote{conservative} relative error bar of $10^{-3}$. Indeed, in their analysis, the no-screening model falls
narrowly out of this range, but other uncertainties might influence the result. In any case, such an important claim should be confirmed independently.

\begin{acks}
I am grateful to the anonymous referee for suggesting the possibility of particle-in-cell simulations for dynamic screening. I am indebted to the late Hugh DeWitt at Lawrence Livermore National Laboratory, with whom I had numerous most inspiring discussions on the topic.
\end{acks}

\end{document}